\documentclass[%
reprint,
superscriptaddress,
showpacs,preprintnumbers,
bibnotes,
 amsmath,amssymb,
 aps,
 prl,
floatfix,
]{revtex4-1}

\usepackage{setspace}
\usepackage{natbib}
\usepackage{graphicx}
\usepackage{dcolumn}
\usepackage{bm}
\usepackage[utf8]{inputenc}
\usepackage[T1]{fontenc}
\usepackage[english]{babel}
\usepackage{amsmath}
\usepackage{amsfonts}
\usepackage{amssymb}
\usepackage{commath}
\usepackage{makeidx}
\usepackage[caption=false]{subfig}
\usepackage{soul}

\DeclareMathOperator\erf{erf}
\DeclareMathOperator{\sech}{sech}

\DeclareGraphicsExtensions{.pdf,.png}
\graphicspath{{./figs_pre/}{./}}
\usepackage[usenames,dvipsnames]{color}   

\usepackage[normalem]{ulem} 

\begin{document}

\title{Functionality of Disorder  in Muscle Mechanics }

\author{Hudson Borja da Rocha}
\email{hudson.borja-da-rocha@polytechnique.edu}
\affiliation{
LMS,  CNRS--UMR  7649,
Ecole Polytechnique, Université Paris-Saclay, 91128 Palaiseau,  France}
\affiliation{
 PMMH, CNRS--UMR 7636 PSL-ESPCI, 10 Rue Vauquelin, 75005 Paris, France}
\author{Lev Truskinovsky}%
\email{lev.truskinovsky@espci.fr}
 \affiliation{
 PMMH, CNRS--UMR 7636 PSL-ESPCI, 10 Rue Vauquelin, 75005 Paris, France}

\date{\today}
\begin{abstract}
A salient feature of skeletal muscles is their ability to take up an applied slack in a microsecond  timescale. Behind this remarkably fast adaptation is a collective folding in a bundle of elastically interacting bistable elements. Since this interaction has long-range character,  the behavior of the system in force and length controlled ensembles is different; in particular, it can have two  distinct order-disorder--type critical points. We show that the account of the disregistry between myosin and actin filaments places the elementary force-producing units of skeletal muscles close to both such critical points.  The ensuing "double-criticality"  contributes to the system's ability to perform robustly  and suggests that the disregistry is functional. 
 \end{abstract}
\maketitle

If an isometrically activated muscle is suddenly shortened, the force first abruptly decreases but then partially recovers over  $\sim$  1 ms  timescale \cite{Podolsky1960, HS71, Irving92}. Behind this remarkably swift contraction is a cooperative conformational change in an assembly of actin-bound myosin heads (cross bridges). Given that this  "power stroke" takes place at a timescale that is much shorter than the timescale of the adenosine triphosphate  (ATP)-driven attachment-detachment  ($\sim$ 100 ms) \cite{howard2001mechanics, Irving02, Higushi_Nature2017}, such fast force recovery is usually interpreted as a \emph{passive} phenomenon \cite{VILFAN_2003, MT10}.  

If it is an applied force, which is controlled, the mean-field theory of fast force recovery, viewing filaments as rigid and cross bridges as parallel \cite{CT_PRE}, predicts metastability associated with a coherent response \cite{CAT_PRL}.  It also predicts the existence of an order-disorder--type critical point, and it was argued that this critical point plays an essential role in the functioning of the muscle machinery \cite{CT_JMPS_2017,CT_ROPP}. This is consistent with the fact that critical systems are ubiquitous in biology because of their adaptive advantages, in particular,   their robustness in the face of random perturbations  \cite{Balleza2008, BegTime2012, Mora2011, Bialek2014, Levine2015}.  Criticality is often linked  to marginal stability and, indeed, skeletal muscles are  known to exhibit near zero passive rigidity in physiological (isometric contractions) conditions   \cite{HS71, Brunello2014, Lombardikappa, Linari19982459}.

\begin{figure}
\includegraphics[width=0.48\textwidth]{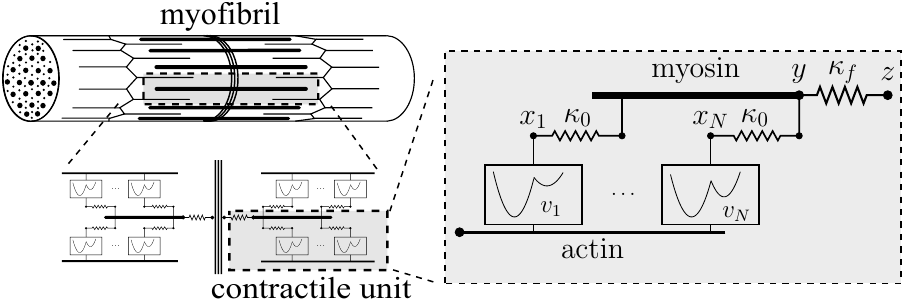}
\caption{ Schematic representation of a muscle myofibril, of an elementary contractile unit (half-sarcomere) and  of a parallel bundle of   $N$ cross bridges.  In the model, the double-well potentials are mimicked by spin variables.}
\label{fig:HSBundle}
\end{figure}

The mechanical functioning of this force generated system system is complicated by the fact that muscle architecture involves both\emph{ parallel} and \emph{series} connections (see Fig.~\ref{fig:HSBundle}). Parallel elements respond to a common displacement (hard device, Helmholtz ensemble), while series structures sense a common force (soft device, Gibbs ensemble). To fold coherently, individual contractile units should be able to coordinate in both types of loading conditions; however,  the dominance of long-range interactions  \cite{Ruffo2009, RuffoPRL} induces \emph{different} collective behavior   in force   and length   controlled   ensembles \cite{CAT_PRL}. In particular, the critical points corresponding to length and force clamp loading conditions are strictly distinct \cite{CT_ROPP}.  

In realistic conditions, however, they turn out to be close to each other and, to ensure the robustness  of the response   under a broad range of mechanical stimuli (flexibility) \cite{Munoz_RevModPhys_2018}, the system can still be poised in the vicinity of \emph{both} critical points.

 In this Letter, we argue that such "double criticality" is actualized in the system of muscle cross bridges due to quenched disorder.  While skeletal muscles are often compared to ideal crystals, the perfect ordering is compromised by the intrinsic disregistry between the periodicities of myosin cross bridges and actin binding sites.  Binding of cross bridges is restricted to incompatibly placed segments on actin filaments (target zones), and experimental studies based on electron microscopy and x-ray diffraction  suggest that myosin heads are bound to actin at seemingly random positions \cite{Tregear98, Tregear20043009}.  To gain an insight into the role of  variable offsets,  we assume that  the  attachment sites  are indeed chosen at random and show that this  gives us an analytically tractable model.  
 
The idea that actomyosin disregistry brings the system's stiffness to zero was pioneered in \cite{HT96}. More recently the utility of quenched disorder for the \emph{active} aspects of muscle mechanics has been  advocated in \cite{Egan12092017}.  The beneficial role of random inhomogeneity has been established in many  other fields of physics from high-temperature superconductivity in electronic materials \cite{Zaanen_Nature_2010}  to  Griffiths phases in brain networks \cite{Munoz_Nature_2013}. 
  
To explore the reachability of the "double criticality" condition in realistic conditions,  we reduce the description of the system of interacting cross bridges to a  random field Ising model (RFIM) and compute the equilibrium free energy applying  techniques from the theory of glassy systems \cite{SGPedestrians}.  We then use the available experimental data on skeletal muscles to justify the claim that quenched disorder is the main factor ensuring the targeted mechanical response.

We associate with each cross bridge a spin variable  $x$  taking the value  $ 0$ in the pre-power-stroke state (unfolded conformation) and $ -1$ in the post-power-stroke state (folded conformation).  Each spin element is then placed in series with a linear elastic spring of stiffness $\kappa_0$.  If we nondimensionalize lengths by the  power-stroke size $a$ and energy by $\kappa_0 a^2$,   the  dimensionless energy of a cross bridge reads
$
(1+ x)v+\frac{1}{2} (y-x)^2, 
$
where $ y$ is the dimensionless displacement of  myosin relative to  actin  and  $ v$ is the dimensionless energetic bias, see Fig.\ref{fig:HSBundle}.   To model disregistry, we assume that the parameter $v$ is different for different cross bridges \footnote{ This form of the quenched disorder is equivalent to the explicit introduction of a pre-strain in each of the linear springs and is also  a signature of spatially inhomogeneous ATP driving}.

Consider now a parallel bundle of $N$ cross bridges shown schematically in Fig.~\ref{fig:HSBundle}. Individual cross bridges are attached to a backbone composed of myosin tails. The elasticity of the backbone can be accounted through a lump spring of stiffness $\kappa_f$  in series with the bundle \cite{LombardiLambdaf, CT_JMPS, PhysRevLett.75.2618}.  The   system   loaded in a hard device is  then  characterized by the  dimensionless  energy
\begin{equation}\label{eq:enebundlehs}
E=\sum_{i=1}^{N}[(1+ x_i)v_i+\frac{1}{2} (y-x_i)^2]+N\frac{\lambda_f}{2}(z-y)^2,
\end{equation} 
where $z$ is the  applied displacement and  $\lambda_f=\kappa_f/(N\kappa_0)$. We  assume that  the parameters $v_i$ are  independent identically distributed random variables with probability density $p(v)$.

If we replace  variables $x_i$ by $s_i=2x_i+1= \pm 1$ and adiabatically eliminate $y$,  assuming that $ \partial E/\partial y=0$, the energy  \eqref{fig:HSBundle}  takes the form
 $$E=- J/(2N) \sum_{i,j} s_i s_j -\sum_i h_i s_i +c,$$ 
 where $J=1/4(1+\lambda_f)$, $c$ is a $z$ dependent  constant, and the coefficients $h_i$ are  linear  in  $v_i$ (see Supplemental Material \cite{SM}). We can then conclude that  \eqref{fig:HSBundle}  is a version  of the mean-field RFIM,  which is explicitly  solvable \cite{SchneiderPytte, Lenka_PRL_2010}.

 Using the  self-averaging  property of the free energy  in the thermodynamic limit, we  write 
 $$\mathcal{F}(\beta,z)=-\lim_{N\rightarrow\infty} (N\beta)^{-1}\langle\log \mathcal{Z}(\beta,z;\{v\})\rangle_v,$$
where the averaging  $\langle  \cdot \rangle_v$ is over the disorder,  $\beta= \kappa_0 a^2/(k_B T)$,  and  
$$
 \mathcal{Z}=\displaystyle \int dy \sum_{x\in\{0,-1\}^N}  \exp ( -\beta E(\boldsymbol{x},y,z;\{v\}). 
$$
 In the thermodynamic limit, we obtain \cite{SM}  
\begin{equation}\label{eq:FEEB}
\begin{split}
\mathcal{F}(\beta,z)&=\frac{\lambda_f}{2}(z-y_0)^2 +\frac{1}{4}(y_0+1)^2+\frac{1}{2}(\frac{y_0^2}{2}+v_0)
\\&-\frac{1}{\beta }\int dv \,p(v)
  \log \left[2 \cosh [\frac{\beta}{4}(1+2y_0-2v)]\right],
\end{split}
\end{equation}
where $y_0 $ must solve the self-consistency equation
\begin{equation}\label{eq:scqd}
y_0=\frac{2\lambda_f z -1}{2(\lambda_f+1)}+\int dv \, \frac{p(v)}{2(\lambda_f+1)}
  \tanh\left[ \frac{\beta}{4}(1-2v+2y_0)\right].
\end{equation}
The multiplicity of solutions of Eq.~\eqref{eq:scqd} is a result of the nonconvexity of the free energy with respect to $y$, which is ultimately an effect of long-range interactions. The multiplicity  leads to the possibility of discontinuous   tension-elongation curves  $t=\partial\mathcal{F}/\partial z =\lambda_f(z-y_0)$.   

\begin{figure}[ht!]
\centering
\includegraphics[width=.48\textwidth]{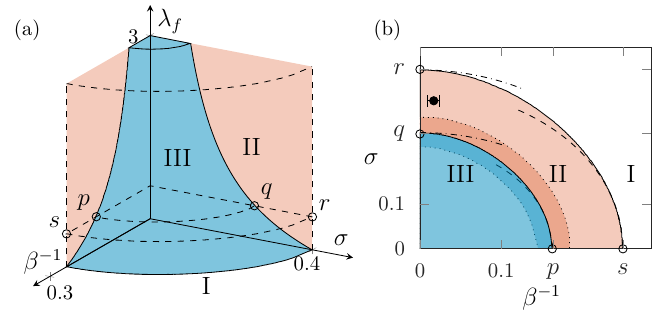}
\caption{ (a) Configuration of phases I, II, and III in the parameter space ($1/\beta$,  $\sigma$ , $\lambda_f$). (b) A  section of this phase diagram corresponding to  $\lambda_f=0.54 \pm 0.2$; the shadowed  region near the boundary of II and III reflects the uncertainty in $\lambda_f$. The realistic dataset for  skeletal muscles is presented in (b) by a  filled circle with the superimposed error bars  indicating uncertainty in temperature.  Analytic approximations in (b): dashed-dotted lines indicate low temperature; dashed lines indicate  low disorder.}
\label{fig:3DPD}
\end{figure}

\begin{figure}
\includegraphics[scale=0.8]{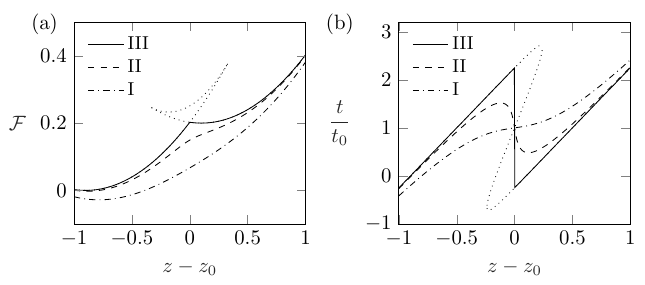}
\caption{ (a) Representative  Helmholtz free energies in each of the phases I, II, III.   (b) The corresponding tension-elongation relations;  $z_0=(1+\lambda_f)v_0 /\lambda_f -1/2$; $t_0=v_0$.}
\label{fig:FETE123}
\end{figure}

If we assume that the disorder is Gaussian 
$p(v)= (2 \pi \sigma^2)^{-1/2}\exp(-\frac{(v-v_0)^2}{2 \sigma^2})$, the behavior of the system will be fully defined by  the temperature $1/\beta$,   the variance of  disorder $\sigma^2$, and  the  parameter $\lambda_f$, characterizing  the degree of elastic coupling.  The  resulting phase  diagram is  shown in Fig.~\ref{fig:3DPD}. The disorder-free section $\sigma=0$  of this diagram was previously studied in \cite{CT_ROPP}. At  $\sigma>0$ the system responds as if it was subjected to a higher effective temperature \cite{Roux_PRE_2000, Politi_PRE_2002}. The  Helmholtz free energy $\mathcal{F}(\beta,z)$ and the tension-elongation relations $t(\beta,z)$ in the three  phases I, II, and III are  illustrated  in Fig.~\ref{fig:FETE123}.

In phase I, the cooperativity is absent and the cross bridges fluctuate independently. In phase III, the cross bridges can  synchronously switch between two  "pure states".  In the intermediate phase II, the tension-elongation relation exhibits negative stiffness.  The boundary between phases II and III is defined by the condition that $\partial^2\tilde{\mathcal{F}}(\beta,z,y)/\partial y^2= 0$,  which is a condition that the three roots of  \eqref{eq:scqd} collapse into one. 

In the limiting case   $\sigma\rightarrow 0$, the point $p$ in Fig.~\ref{fig:3DPD}(b)  is at  $\beta=4(\lambda_f+1)$. Around this point, the $p-q$ curve is described accurately by the low-disorder approximation $\beta_{e}=4(\lambda_f+1)$ where 
 $\beta_{e}= (\beta^{-2}+\sigma^2/2)^{-1/2}$
 is the  inverse\emph{ effective temperature} (see Supplemental Material \cite{SM}).  In another limiting case $\beta\rightarrow \infty$,  the point  $q$ can be found from the equation $  \sigma =1/\sqrt{2\pi  (\lambda_f+1)}$ and around this point the $p-q$ curve is given by the small temperature approximation $  \sigma_e =1/\sqrt{2\pi  (\lambda_f+1)}$, where
  $\sigma_e^2=(\sigma^2+2\beta^{-2})= 2\beta_{e}^{-2}$
is the  variance of the \emph{effective disorder} \cite{SM}. 
 
\begin{figure}
\includegraphics[scale=.8 ]{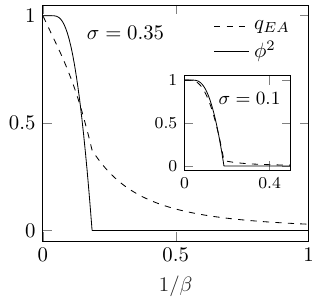}
\caption{The behavior of the  parameter $\phi^2$  (solid lines)  and the Edwards-Anderson parameter $q_{\text{EA}}$ (dashed lines)  near the boundary between phases  II and III    at the realistic value of disorder. (Inset)  The case of weak disorder.}
\label{fig:q_m-HD}
\end{figure}

 The boundary between phases  II and III  marks a  second-order phase transition:   the  order parameter   $\phi= N^{-1}\sum_{i=1}^{N}  \left<s_i\right >_{\beta}$, where    $ \langle \cdot \rangle_{\beta}$ is the thermal average,   is double valued  in  phase III  and  single valued in  phase II. To distinguish between different microscopic configurations,  we also compute the Edwards-Anderson (overlap)  parameter $q_{\text{EA}}=N^{-1}\sum_{i=1}^{N} \langle \langle s_i\rangle^{2}_{\beta} \rangle_v$  \cite{SM}. If   $q_{\text{EA}}\neq 0$   while  $\phi=0$, the pre- and post-power-stroke  symmetry is   broken and cross bridges may be locally frozen in either of the two states, even though such local ordering in time does not imply any spatial order.    Figure~\ref{fig:q_m-HD} shows that   $q_{\text{EA}}$ is indeed different from zero in the  phase II close to the $p-q$ boundary, which indicates  \emph{weakly glassy}  behavior \cite{SchneiderPytte, Vilfan_1987, Lenka_PRL_2010}. This is a hint  that, in a more realistic model, where the finite backbone stiffness is taken into account,  a real "strain glass"  phase \cite{Otsuka_PRL_2006, Lookman_PRB_2012} is likely to appear. 
   
To find the boundary between phases I and II,  we need to solve the equation $\partial^2 \mathcal{F}/\partial z^2 = 0$ or $ \partial y_0/\partial z=1$, where   $y_0$  is a solution of  \eqref{eq:scqd}.  When  $\sigma=0$, we obtain  $\beta=4$, which defines the location of point $s$ in Fig.~\ref{fig:3DPD}(b) (see also \cite{CAT_PRL,CT_JMPS}). The low-disorder approximation gives $\beta_{e}=4$.  In another limiting case $\beta\rightarrow \infty$, the location of the point $r$ in Fig.~\ref{fig:3DPD}(b) is  given by  $\sigma=\sigma_e=\sqrt{1/ 2\pi}$.

\begin{figure}[ht]
\centering
\includegraphics[scale=.8]{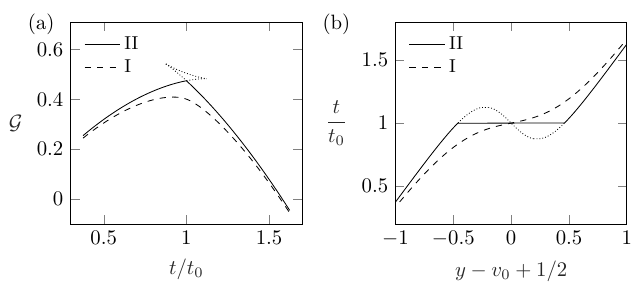}
\caption{(a) Representative Gibbs free energies in each of the phases I and II.    (b) The corresponding tension-elongation curves; $z_0=(1+\lambda_f)v_0 /\lambda_f -1/2$; $t_0=v_0$.  }
\label{fig:TensionElongationSD}
\end{figure}

The boundary   between the phases I and II  can be also  interpreted as a line of second-order phase transitions, but now in the soft device (force clamp) ensemble. In this case, the presence of a series spring  is irrelevant and we can   assume that  $\lambda_f\rightarrow 0$,  $z\rightarrow \infty$, but  $\lambda_f z \rightarrow t $, where  tension $t$ is the new control parameter.  Following the  approach used  in the case of a hard device, we  similarly obtain the Gibbs free energy
$
\mathcal{G}(\beta,t)$ and compute  
the tension-elongation relation  $y=-  \partial \mathcal{G}/\partial t$, (see Supplemental Material \cite{SM}).

In  Fig.~\ref{fig:TensionElongationSD},  we show that the soft device tension-elongation relation in phase II is monotone but discontinuous. On the boundary of I and II [see  Fig.~\ref{fig:3DPD}(b)],  the stiffness becomes zero in stall conditions, which means that it is a   set of critical points in the soft device ensemble. This line, targeted numerically in  \cite{HT96}, represents regimes that can be expected to deliver the optimal trade-off between robustness and flexibility in the soft device \cite{kauffman1993origins, Darabos2011}.  

So far,   we have operated under an implicit assumption that in the thermodynamic limit   $\kappa_f \to \infty$, while  $\lambda_f$  remains finite. This assumption is based on the picture of a myosin filament as a parallel arrangement of $N$  myosin tails, all contributing to the lump stiffness of the backbone.  An alternative assumption may be that the effective stiffness of the backbone $\kappa_f$  does not depend  on the number of attached cross bridges $N$ and, in this case, we have a different scaling $\lambda_f\sim N^{-1}$. Then   Fig.~\ref{fig:3DPD}(a),  illustrating the size effect,  suggests that the quasicritical behavior should be tightly linked to the  particular (optimal)  number of cross bridges.

To apply our  results  to  a realistic muscle system, we use the data for \textit{Rana temporaria}  at $T=277.15$ K  \cite{CT_ROPP}. From structural analysis, we obtain the value $a \sim 10 $ nm \cite{Dominguez_Cell,Rayment50,Rayment58}.  Measurements of the fiber stiffness  in \textit{rigor mortis},   where all the 294 cross bridges per half-sarcomere were attached, produced the estimate  $\kappa_0=2.7\pm 0.9$ pN/nm \cite{Brunello2014,Lombardikappa}. The number of attached cross bridges in physiological conditions is $N =106 \pm 11$ and  experimental measurements at different  $N$  converge on the value $\kappa_f=154 \pm 8$ pN.nm$^{-1}$ for the lump filaments stiffness  \cite{WAKABAYASHI19942422lf,HUXLEY19942411lf,Irving02}. This gives $\lambda_f=0.54 \pm 0.2 $.  Knowing $\kappa_0$ and $a$ we can estimate the nondimensional inverse temperature to be $\beta=71 \pm 26$.   
  
Now,  for $y>y_*$, where $y_*=v_0-1/2$, the ground state of a single cross bridge is in the pre-power-stroke state,  while for $y<y_*$ it is in the post-power-stroke state, so    $y_*$  represents the characteristic offset for an individual cross bridge. Knowing that $y_*\sim4$nm \cite{HS71,HT96}, we conclude that $v_0\sim 24.3$pN $/(\kappa_0 a)$. It was experimentally shown in \cite{Tregear20043009} that at least 60$\%$ of the cross bridges are axially displaced within half of the spacing between actin monomers, which corresponds to $\sim 2.76$ nm shift from the nearest actin binding site (see also  \cite{HT96}). Given the linear relation between $v_0$ and  $y_*$ with the proportionality coefficient equal to one, the variances of these two quantities are the same. If the axial offsets are Gaussian random numbers, we can estimate the standard deviation of the energetic bias $\sigma\sim 3.3$nm$/a$ (see Supplemental Material \cite{SM}).
 
\begin{figure}[ht]
\centering
\includegraphics[width=.48\textwidth]{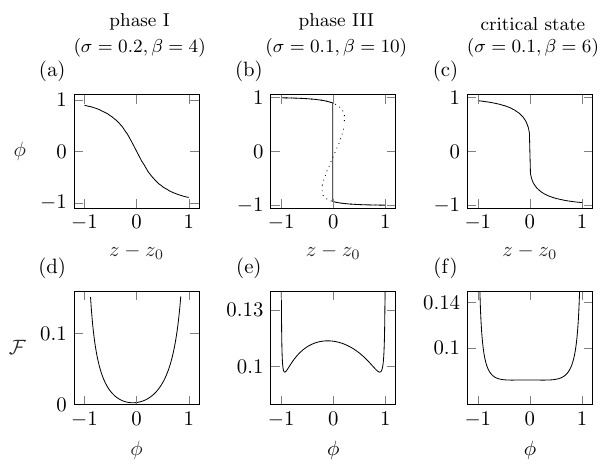}
\caption{The structure of the energy barriers in different regimes for the case of the hard device. (a)--(c)  $z$ dependence of  the order parameter $\phi= N^{-1}\sum_{i=1}^{N}  \langle s_i\rangle$ in  different regimes; (d)--(f)   matching  free energies  at fixed $z=z_0$. $\lambda_f=0.35$ and $v_0=0.1$}
\label{fig:FEOP}
\end{figure}

 Based on these data we find that, rather remarkably,  the system appears to be operating in a narrow domain of stability of phase II, close to both critical lines  $p-q$ and $r-s$ [see the point marked by a  filled circle in Fig.~\ref{fig:3DPD}(b)].  The gap between these boundaries corresponds to $\sim 1$ nm difference in the cross bridge attachment positions, which is rather small given that the size of a single actin monomer is about $5.5$ nm.  The mechanical responses in the adjacent critical regimes  are structurally similar; however, if in the hard device ensemble we can expect coherent fluctuations of \emph{stress}   (infinite rigidity), in the soft device, criticality would manifest itself through system size correlations of \emph{strain} (zero rigidity).

The special nature of the critical regimes is illustrated in Fig.~\ref{fig:FEOP} for the case of a hard device.  In phase I, the response is uncorrelated, and the collective power stroke is impossible [Figs.~\ref{fig:FEOP}(a) and 6(d)].  In phase III, the response is synchronous but at the cost of crossing an energetic barrier that diverges in the thermodynamic limit ($\mathcal{F}$ is the free energy \emph{per} cross bridge), which facilitates freezing in the pure states, [see Figs.~\ref{fig:FEOP}(b) and 6(e)]. The advantage of the critical regime is that the system can perform the \emph{collective} stroke without crossing a prohibitively high macroscopic barrier, [see Figs.~\ref{fig:FEOP}(c) and 6(f)]. The analysis is similar for the case of a soft device.
 
Our study then suggests that evolution might have used quenched disorder to tune the muscle machinery to perform near the conditions where both the Helmholtz and the Gibbs free energies are singular. Such design is highly functional when elementary force-producing units  are   loaded in a mixed, soft-hard device.  We recall that the muscle architecture is characterized by hierarchical structures with coupled modular elements loaded both in parallel and in series. In such systems, the proximity to only one of the two critical points will not be sufficient to ensure high performance in a broad range of conditions \cite{Munoz_RevModPhys_2018, Bialek_ROPP_2018}. Moreover, as we show in the Supplemental Material \cite{SM},   the very idea of ensemble independent  \emph{local} constitutive relations for such systems becomes questionable.

 In conclusion, we established new links between muscle physiology and the theory of spin glasses and revealed a  tight relation between actomyosin disregistry and the optimal mechanical performance of the  force-generating machinery. At a price of  neglecting   many important features of actual muscles, we were   able to focus attention on the   role of quenched disorder in the functioning of this biological system. The observed  \emph{glassiness} in the regime of isometric contractions allows the system to  access the whole spectrum of rigidities from zero (adaptability, fluidity) to infinite  (control, solidity) and may serve as the factor ensuring the largest dynamic repertoire of the "muscle material".  Similar disorder-mediated tuning towards criticality can be expected in other biological systems relying on bistability and long-range interactions \citep{CT_PRE}, including hair cells, which employ elastically coupled  gating springs  \cite{Bormuth20052014} and focal adhesions with their cell adhesion molecules bound to a common substrate \cite{Safran_RevModPhys.85.1327}.

\begin{acknowledgments}
The authors thank M. Caruel and R. Garcia-Garcia for helpful discussions. H.B.R. received support from an Ecole Polytechnique Fellowship; L. T. was supported by Grant No. ANR-10-IDEX-0001-02 PSL.
\end{acknowledgments}

\bibliographystyle{apsrev4-1}

\appendix

\section{Supplemental Material for the paper "Functionality of Disorder in Muscle Mechanics" }

\subsubsection{Mapping to the Random-Field Ising Model}

We start with the energy function  (1) in the main text  and assume that the internal variable $y$ is eliminated  using the condition $ \partial E/\partial y=0$. Then,
\begin{equation} \nonumber
y=\frac{\lambda_f z}{1+\lambda_f}+\frac{1}{N(1+\lambda_f)}\sum_i x_i.
\end{equation}
and the relaxed energy reads
\begin{equation}\label{eq:enerfim1}
\begin{split} \nonumber
E(x_i,z)&=-\frac{1}{2 N (1+\lambda_f)}\left(\sum_ix_i\right)^2\\+\sum_i (1 &+x_i)v_i  - \frac{\lambda_f z}{1+\lambda_f} \sum_i x_i+\sum_i \frac{x_i^2}{2}+ \frac{N\lambda_fz^2}{2(1+\lambda_f)}
\end{split}
\end{equation}
Since $x_i$ is either 0 or -1, we may write $\sum_i x_i^2=-\sum_i x_i$ and  $\left(\sum_ix_i\right)^2=\sum_i \sum_j x_i x_j=\sum_{i,j} x_i x_j$. In terms of  spin variables, $2x_i=s_i-1,\, \text{with }  s_i=\pm 1
$ the relaxed energy can be written as,
\begin{equation} 
\begin{split}
E(s_i,z)&=-\frac{1}{8 N (1+\lambda_f)}\sum_{i,j} s_i s_j  \\-& \sum_i\left(\frac{2\lambda_f z-1}{4(1+\lambda_f)}+\frac{1}{4}-\frac{v_i}{2}\right) s_i \\
+&\sum_i \left(\frac{\lambda_f z(1+z)}{2(1+\lambda_f)}+\frac{1}{4}+\frac{v_i}{2}-\frac{1}{8(1+\lambda_f)}\right)\\&=-\frac{J}{2N} \sum_{i,j} s_i s_j -\sum_i h_i s_i +f(z).
\end{split}
\end{equation}
where $J=\frac{1}{4(1+\lambda_f)}$, $h_i=\frac{2\lambda_f z-1}{4(1+\lambda_f)}+\frac{1}{4}-\frac{v_i}{2}$ and $f(z)=\sum_i \frac{\lambda_f z(1+z)}{2(1+\lambda_f)}+\frac{1}{4}+\frac{v_i}{2}-\frac{1}{8(1+\lambda_f)}$.

\subsubsection{Computation of the free energy}

 Using the  self-averaging  property of the free energy  in the thermodynamic limit, we  write 
 $$\mathcal{F}(\beta,z)=-\lim_{N\rightarrow\infty} (N\beta)^{-1}\langle\log \mathcal{Z}(\beta,z;\{v\})\rangle_v,$$
where the averaging  $\langle\cdot\rangle_v$ is over the disorder,  $\beta= \kappa_0 a^2/(k_B T)$,  and  
 \begin{equation} \nonumber
 \mathcal{Z}=\displaystyle \int dy \sum_{x\in\{0,-1\}^N}  \exp ( -\beta E(\boldsymbol{x},y,z;\{v\}).
 \end{equation}
 The mean field nature of the model allows one to rewrite this expression  in the form  
\begin{equation} \nonumber
\mathcal{Z} =  \int dy \exp ( -\beta N [ \frac{\lambda_f}{2}(z-y)^2 -\frac{1}{\beta N} \sum_{i=1}^{N}\log \tilde{\mathcal{Z}}] ),  
\end{equation} 
 where  $\tilde{\mathcal{Z}}=e^{-\frac{\beta}{2}(y+1)^2 }+e^{-\beta ( y^2/2+v_i)}$  is the   partition function of  a single Huxley-Simmons element \cite{HS71SM, CT_PRESM}. In the thermodynamic limit, we can use the   saddle-point approximation to obtain 
$\mathcal{F}(\beta,z)=\tilde{\mathcal{F}}(y_0,\beta,z),$ where 
$\tilde{\mathcal{F}}(y,\beta,z)=\beta \frac{\lambda_f}{2}(z-y)^2 -  \langle\log \tilde{\mathcal{Z}}\rangle_v$ and $y_0(\beta,z)$ is the minimum of $\tilde{\mathcal{F}}$. More explicitly, 
\begin{equation}\label{eq:FE}
\begin{split}
\mathcal{F}(\beta,z)&=\frac{\lambda_f}{2}(z-y_0)^2 +\frac{1}{4}(y_0+1)^2+\frac{1}{2}(\frac{y_0^2}{2}+v_0)
\\&-\frac{1}{\beta }\int dv \,p(v)
  \log \left[2 \cosh [\frac{\beta}{4}(1+2y_0-2v)]\right],
\end{split}
\end{equation}
where $y_0 $ solves the self-consistency equation,
\begin{equation}\nonumber
y_0=\frac{2\lambda_f z -1}{2(\lambda_f+1)}+\int dv \, \frac{p(v)}{2(\lambda_f+1)}
  \tanh\left[ \frac{\beta}{4}(1-2v+2y_0)\right].
\end{equation}

\subsubsection{Boundary between phases II and III}

Using the expression for the  partial free energy,
\begin{equation}\nonumber
\begin{split}
\tilde{\mathcal{F}}(\beta,z,y)&=\frac{\lambda_f}{2}(z-y)^2 +\frac{1}{4}(y+1)^2+\frac{1}{2}(y^2/2+v_0)
\\&-\frac{1}{\beta }\int dv \,p(v)
  \log \left[2 \cosh [\frac{\beta}{4}(1+2y-2v)]\right]
\end{split}
\end{equation}
we can write the condition $\partial^2\tilde{\mathcal{F}}(\beta,z,y)/\partial y^2= 0$ in the form 
\begin{equation}\nonumber
\lambda_f+1-\frac{\beta}{4}\int dv\,p(v)\sech^2\frac{\beta}{4}(1-2v+2y_0)=0.
\end{equation}
If we use the Gaussian distribution of disorder introduced in the main text and use new variables  $\eta=\beta(1+2y_0)/2$ and $\bar{v}=\beta v$ we can rewrite this equation in the form
\begin{equation} \nonumber
\lambda_f+1-\frac{\beta}{4}\int d\bar{v}\frac{e^{-\frac{(\bar{v}-\beta v_0)^2}{2\sigma^2\beta^2}}}{\sqrt{2\pi \sigma^2 \beta^2}} \sech^2\frac{1}{2}(\eta-\bar{v})=0.
\end{equation}
Note that the variance of disorder appears in this formula only in the combination 
$\sigma^2 \beta^2$. This means that,  modulo some obvious adjustments,
the small disorder $\sigma \to 0$  and  large temperature   $\beta \to 0$ limits are complimentary.  The same can be said about the  small temperature $\beta \to \infty$ and the large disorder $\sigma \to \infty$ limits.  

\emph{ Zero disorder limit.} In  the limit $\sigma\rightarrow 0$  we have  $p(v)\rightarrow \delta(v-v_0)$ and   the boundary between phase II and III is defined by the equation
\begin{equation}\nonumber
\lambda_f+1=\frac{\beta}{4}\sech^2\frac{\beta}{4}(1-2v_0+2y_0).
\end{equation}
Since $\sech^2 x \in [0,1]$, this equation does not have  solutions $y_0$ for  $\beta>4(\lambda_f+1)$ and therefore the point $r$ is defined by the condition $\beta=4(\lambda_f+1)$.

 To get the next term of the asymptotic expansion we  introduce the new variable $\xi=(1-2v+2y_0)/4$ and  assume that the temperature is large $\beta \to 0$. Then we can expand 
$
\log \sech^2 \beta \xi \approx -\beta^2\xi^2+O(\beta^4),
$
which implies that $\sech^2 \beta\xi\approx e^{-\beta^2\xi^2}$.  Using this approximation we can compute the integral and represent  the boundary  between phase II and III   in the form
\begin{equation}\nonumber
\lambda_f+1=\frac{e^{\frac{-(y_0- v_0+1/2)^2}{2( 2T^2+ \sigma ^2)}}}{2\sqrt{2(2T^2+\sigma^2)}}.
\end{equation}
where $T=1/\beta$. Since $e^{-x^2}\in (0,1]$ the criticality condition is
\begin{equation} \nonumber
(\lambda_f+1)2\sqrt{2(2T^2+\sigma^2)}=1
\end{equation}
 The equivalent quenched disorder is then defined by the condition $\sigma^2_{eq}=2T^2+\sigma^2$.

\emph{Zero temperature limit. }In the zero temperature limit $\beta\rightarrow \infty$ we use the fact that 
$
\lim_{k\rightarrow \infty} \frac{k}{2} \sech^2 kx \rightarrow \delta(x).
$
to rewrite the equation defining the  boundary between phase II and III  in the form 
\begin{equation}\nonumber
(\lambda_f+1)\sqrt{2\pi \sigma^2}= e^{-\frac{(y_0+1/2-v_0)^2}{2\sigma^2}}.
\end{equation}
Here the \textit{r.h.s} is defined in the interval $(0,1]$ and therefore there are no solutions $y_0$ if $(\lambda_f+1)\sqrt{2\pi \sigma^2}>1$  where we used the fact that  $\sigma, \lambda_f>0$. The  point $q$  is then defined by the condition  $(\lambda_f+1)\sqrt{2\pi \sigma^2}=1$. 

To obtain the next term of the asymptotic expansion we assume that disorder is large  $\sigma\rightarrow \infty$. In this case we can still approximate the function $\sech^2(x)$  by the   Gaussian distribution but now the approximation should be good not at $x=0$ but globally. To this end we need to  require that the two functions are equally normalized
\begin{equation} \nonumber
\begin{split}
\frac{1}{4T}\int dv  \sech^2 & \frac{1-2v-y_0}{4T}  \\ 
&=\frac{1}{\sqrt{4\pi T^2}}\int dv e^{-\frac{(v-y_0-1/2)^2}{4T^2}} = 1,
\end{split}
\end{equation}
where again $T=1/\beta$. With this normalization the integral can be again computed and we obtain the condition
\begin{equation} \nonumber
(\lambda_f+1)\sqrt{2\pi}=\frac{e^{\frac{-(y- v_0+1/2)^2}{2( 2T^2+ \sigma ^2)}}}{\sqrt{2T^2+\sigma^2}}.
\end{equation}
The criticality criterion is then 
\begin{equation} \nonumber
(\lambda_f+1)\sqrt{2\pi(2T^2+\sigma^2)}=1,
\end{equation}
which allows us to introduce the effective disorder by the condition $\sigma_e^2=2T^2+\sigma^2$.

\subsubsection{ Gibbs free energy}

 In the case of soft device the  relevant potential is,
\begin{equation}
G =\sum_{i=1}^{N}\left[(1+ x_i)v_i+\frac{1}{2} (y-x_i)^2\right]-t y.
\end{equation} 
Following the  approach used  in the case of hard device, we obtain the expression for the Gibbs free energy
\begin{equation}\label{eq:Gibs}
\begin{split}
\mathcal{G}(\beta,t)&=-ty_0 +\frac{1}{4}(y_0+1)^2+\frac{1}{2}(\frac{y_0^2}{2}+v_0)
\\&-\frac{1}{\beta }\int dv \,p(v)
  \log \left[2 \cosh [\frac{\beta}{4}(1+2y_0-2v)]\right]
\end{split}
\end{equation}
where   now $y_0$  solves the equation
\begin{equation}\label{eq:scsd}
t=y_0+\frac{1}{2}-\frac{1}{2}\int dv \, p(v)
  \tanh\left[ \frac{\beta}{4}(1-2v+2y_0)\right].
\end{equation}
The tension elongation relation is then a solution of $y=-  \partial \mathcal{G}/\partial t$.
 
\subsubsection{ Edwards-Anderson order parameter}
 
In the absence of disorder, a natural  order parameter is
\begin{equation}\nonumber
\phi=\frac{1}{N}\sum_{i=1}^{N} \left<s_i\right>_T,
\end{equation}
where $s_i=2x_i+1$. To find $\phi(z,\beta)$ we notice that  since all cross-bridges are the same we can write $\phi=2\left<x_i\right>_T+1$ where
\begin{equation} \nonumber
\left<x_i\right>_T=-Z(\beta,z)^{-1} e^{-\beta E(x_i=-1,y_0,z)}
\end{equation}
with $$Z(\beta,z)=e^{-\beta N\left[ \frac{\lambda_f}{2}(z-y_0)^2 -\frac{1}{\beta } \log (e^{-\frac{\beta}{2}(y_0+1)^2 }+e^{-\beta ( y_0^2/2+v)})\right]}.$$ By combining these expressions we obtain
$$\left<x_i\right>_T=-\frac{1}{1+e^{\beta(y_0-v+1/2)}}.$$

In the presence of disorder,  the average values $\left<x_i\right>_T $ are  different for  different cross-bridges and the  macroscopic parameter $\phi(z,\beta)$ is no longer  sufficient to differentiate between microscopic configurations. To this end we can introduce an analogue of the Edwards-Anderson parameter from the  theory of spin glasses 
\begin{equation} \nonumber
q_{EA}=\frac{1}{N}\sum_{i=1}^{N}\left<\left<s_i\right>^{2}_T\right>_v.
\end{equation}
where we distinguish between the thermal average $\left< \cdot \right>_T$ and the ensemble average $\left<A\right>_v=\int dv p(v) A(v)$. If the parameter $\phi$ characterizes the average occupancy of the pre-power stroke state, the nonzero value of  $q_{EA}$ means that individual cross bridges are 'frozen' either in pre- or post-power-stroke states  even if in average,  both states appear to be equally occupied. The knowledge of this parameter is needed, for instance, if one is interested in computing the effect of the random field on mechanical susceptibility (stiffness) \cite{Vilfan_1985SM}

In terms of  the variables $x_i$  the definition of $q_{EA}$ reads
\begin{equation} \nonumber
q_{EA}=\frac{1}{N}\sum_{i=1}^{N}\left[4 \left<\left<x_i\right>^{2}_T\right>_v+4 \left<\left<x_i\right>_T\right>_v+1\right],
\end{equation}
where
\begin{equation} \nonumber
 \left<\left<x_i\right>_T^2\right>_v =\int dv \frac{p(v)}{(1+e^{\beta(y_0-v+1/2)})^2},
\end{equation}
and
\begin{equation} \nonumber
 \left<\left<x_i\right>_T\right>_v =-\int dv \frac{p(v)}{1+e^{\beta(y_0-v+1/2)}}.
\end{equation}

\subsubsection{Boundary between phases I and II}

Note first that 
$
 \frac{\partial t}{\partial z}=\lambda_f(1-\frac{\partial y_0}{\partial z}) ,
$
and therefore to get zero stiffness we must have  
$
 \partial y_0/\partial z=1,
$
Here $y_0$ is found from the  self-consistency condition given by Eq. 5 in the main text  and therefore
\begin{equation}
\begin{split}
\frac{\partial y_0}{\partial z}&=\frac{\lambda_f}{\lambda_f+1} \\&+\frac{\beta}{4(1+\lambda_f)}\int dv \, p(v)\sech^2\left[\frac{\beta}{4}(1-2v+2y_0)\right]\frac{\partial y_0}{\partial z},
\end{split}
\end{equation}
which is equivalent to  
\begin{equation}\nonumber
1=\frac{\beta}{4}\int dv \, p(v)\sech^2\left[\frac{\beta}{4}(1-2v+2y_0)\right].
\end{equation}
The condition that this equation has a root $y_0$ does not contain  $\lambda_f$ and therefore the boundary between phases I and II is $\lambda_f$ independent.

\emph{Zero disorder limit.} In the limit  $\sigma \to 0$ we can again assume that  the probability density $p(v)$ is infinitely localized  and compute the integral explicitly. We  obtain
\begin{equation}\nonumber
\frac{4}{\beta}=\sech^2\frac{\beta}{4}(1-2v+2y_0).
\end{equation}
Since $\sech^2 x \in [0,1]$, this equation does not  have solutions $y_0$ if $\beta<4$, hence $\beta_c=4$, which is the coordinate of our  point $s$. The higher order asymptotic expansion can be obtained following the same procedure as in the case of the boundary between phases II and III.

\emph{Zero temperature limit. }In  the limit $\beta\rightarrow \infty$, we can again use the fact that the function $\frac{k}{2} \sech^2 kx$ converges to the delta function as $k\rightarrow \infty$. Therefore, assuming that  the probability distribution $p(v)$ is  Gaussian we obtain,
\begin{equation}\nonumber
1=\frac{1}{\sqrt{2 \pi \sigma^2}}e^{-\frac{(y_0+1/2-v_0)^2}{2 \sigma^2}}.
\end{equation}
Using the same  arguments as in the zero disorder limit  and noticing that  $e^{-x^2}\in (0,1]$, we conclude that this equation has solution only if $\sigma \geq 1/\sqrt{2\pi}$. Therefore,  the critical value of the disorder in this limit  is $\sigma_c=1/\sqrt{2\pi}$, which corresponds to our point $r$.  The expansion around this point can be obtained as in  the case of the boundary between phases II and III considered above.

\subsubsection{ Axial offset }

\begin{figure}[ht]
\centering
\includegraphics[scale=.7]{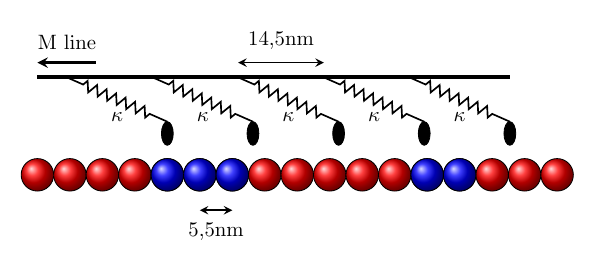}
\caption{Schematic representation of the attachment sites. Each sphere represents an actin monomer; blue color delineate target zones.}
\label{fig:Sarcomere}
\end{figure}

Experimental studies using electron microscopy (EM) and x-ray diffraction have shown that the biding of cross-bridges is restricted to limited segments of the actin filament known as target zones \cite{Tregear98SM,Suzuki2011SM}. These zones   are represented by two to three actin monomers, see Fig.~\ref{fig:Sarcomere}. Moreover, it was found  \cite{Tregear20043009SM}  that the  probability distribution of axial offsets from the target zone center is approximately  Gaussian  and that at least 60$\%$ of the attached cross-bridges  are displaced  within half of the spacing between actin monomers which corresponds to the offset of 2.76nm. 

The offset   can be represented by the reference elongation $y_0=v_0-1/2$  which marks the boundary between pre and post-power stroke states. Because the parameters  $v_0$ and $y_0$ differ by a constant,  the  variance of  $\delta v_0$ is equal  to the variance $\delta y_0$. Hence, placing   disorder in the energetic bias $v_0$ is equivalent to introducing  variable axial offset.

\emph{Gaussian distribution of offsets.} If we suppose that the distribution of axial offsets between the myosin head and the actin biding site is Gaussian we can estimate its standard deviation by noting that the probability that the variable deviation lies in the range $\pm k \sigma$ is given by, 
\begin{equation}
\Pr(\mu-k \sigma \leq X \leq \mu +k \sigma) = \erf(\frac{k}{\sqrt{2}}),
\end{equation}
we then use the fact that 60$\%$ is in the range $\pm$2.76nm to find $k=0.842$ and $\sigma=3.3$nm.

\subsubsection{Critical response in soft and hard ensembles }

\begin{figure}[ht]
\centering
\includegraphics[width=.45\textwidth]{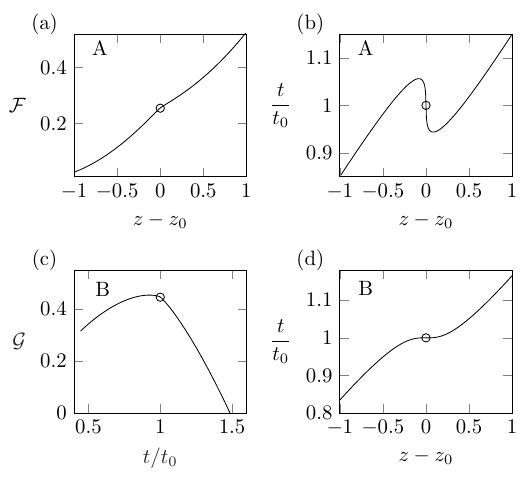}
\caption{The  response of the  system in the critical regimes $A$ and $B$ shown in Fig. 2 of the main text :  (a) and (b) are the Helmholtz free energy and the tension-elongation curve in the hard device ensemble; (c) and (d) are the Gibbs free energy and the associated tension-elongation curve in the soft device ensemble. Critical points are marked by the small circles. }
\label{fig:FreeEnergyStressStrainAB}
\end{figure}

In Fig.~\ref{fig:FreeEnergyStressStrainAB} we illustrate the mechanical responses in the  adjacent critical regimes marked as  $A$ and $B$ in  Fig. 2 of the main text.  In the associated critical points, indicated here  by small circles and intended to represent the physiological regime of isometric contractions,  the susceptibilities diverge. The closeness of these two regimes in the parameter space allows the system to exhibit the whole repertoire of behaviors  from  zero to  infinite rigidity.

\subsubsection{Two  half-sarcomeres in series}

Here we present an elementary illustration of the fact that the equilibrium response of a bundle of  contractile units connected in series and placed in a hard device,  cannot be described by local equilibrium constitutive relations obtained in either soft or hard device ensembles. Instead, the system exhibits an intermediate behavior.

\begin{figure}[ht]
\centering
\includegraphics[scale=.65]{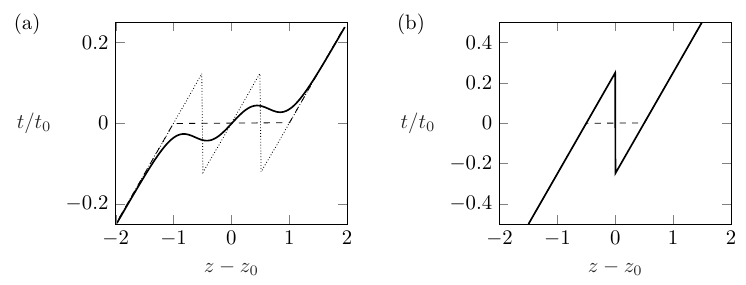}
\caption{(a) Tension elongation relations for a system containing   two half-sarcomeres in series placed in a hard device. Thick line: equilibrium response. Dotted (dashed) line: the response of two contractile elements  in series, each one endowed with its own equilibrium the hard (soft) device constitutive law. (b) Response of a single half-sarcomere. Thick line:  hard device; dashed line:  soft device.  $\beta=30$, $\sigma=0$, $v_0=0$, $\lambda_f=1$.}
\label{fig:series_connection}
\end{figure}

Consider two elementary contractile units in series,  see  \cite{CT_ROPPSM} for the analysis of $M$ such elements. Each of the two elements  represents a parallel connection of $N$ cross-bridges.  The total  energy per cross bridge in dimensionless form  for a system placed in a hard device reads

\begin{equation}
\begin{split}
E_2=\frac{1}{2}\left\lbrace\frac{1}{N}\sum_i^N [(1+ x_{i1})v_{i1}+\frac{1}{2} (y_1-x_{i1})^2+\frac{\lambda_f}{2}(z_1-y_1)^2]\right. \\
+\left.\frac{1}{N}\sum_i^N [(1+ x_{i2})v_{i2}+\frac{1}{2} (y_2-x_{i2})^2+\frac{\lambda_f}{2}(z_2-y_2)^2]\right\rbrace 
\end{split}
\end{equation}
The equilibrium response of the system is obtained by computing the partition function
\begin{equation*}
\mathcal{Z}_2(z,\beta)=\int \exp[-2\beta N E_2]\delta(z_1+z_2-2z)d\boldsymbol{x}
\end{equation*}
where $d\boldsymbol{x}=\prod_{i}^N dx_{i1}dy_1\prod_{j}^N dx_{j2}dy_2$ and $z$ is the (average) elongation imposed on the system. We can rewrite the expression for $\mathcal{Z}_2$ in the form 
\begin{equation}\begin{split}
\mathcal{Z}_2(z,\beta)=\int dy_1dy_2\exp\left\lbrace-\beta N[-\frac{\lambda_f}{2}(z-y_1-y_2)^2\right. \\ 
 \left.-\frac{1}{\beta}\int dv p(v) \log\tilde{\mathcal{Z}}_1(y_1,v)\tilde{\mathcal{Z}}_2(y_2,v)  ] \right\rbrace
 \end{split}
\end{equation}
where 
$\tilde{\mathcal{Z}}_i(y_i,v)=e^{-\frac{\beta}{2}(y_i+1)^2 }+e^{-\beta ( y_i^2/2+v)}$.
The free energy per cross-bridge is then $\mathcal{F}_2(z,\beta)=-\frac{1}{2N}\log\mathcal{Z}_2(z,\beta)$. The equilibrium tension-elongation relation for this system,  obtained from the relation $t(z,\beta)=\partial \mathcal{F}_2(z,\beta)/\partial z$,  is shown by  the thick line  in Fig.~\ref{fig:series_connection}(a). Similar thick line in Fig.~\ref{fig:series_connection}(b) shows the equilibrium response of a single contractile element placed in the hard device.

We now compare this behavior with the one obtained under the assumption that the two elements in series are characterized by their equilibrium free energies computed either in a hard or in a  soft ensembles. 

For instance,  using the  hard device ensemble  we can write  the total (Helmholtz) free energy of the two element system in the form $E_2^{hd}=\mathcal{F}(z_1,\beta)+\mathcal{F}(z-z_1,\beta)$, where $\mathcal{F}$ is the free energy of a half-sarcomere given by Eq.~\ref{eq:FE}. The extra variable $z_1$ can be eliminated using the equilibrium condition $\partial \mathcal{F}(z_1,\beta)/\partial z_1=\partial \mathcal{F}(z-z_1,\beta)/\partial z_1$. The resulting tension elongation curve  is shown in Fig.~\ref{fig:series_connection} (a)  by a dotted line. 

 Similar analysis can be performed  based on the response functions for the elements loaded in  a soft device.  Here we need to use equilibrium (Gibbs) free energies of the elements (Eq. 5 in the main text) and  since the elements in series  share the value of tension we obtain $G_2^{SD}=2\mathcal{G}(t,\beta)$.  The ensuing response of the series bundle is shown in Fig.~\ref{fig:series_connection}(a) by a dashed line. In Fig.~\ref{fig:series_connection}(b), the dashed line show the equilibrium response of a single contractile element loaded in a soft device.

Observe, first,   that the equilibrium response  predicted by the two 'constitutive models'  contains  discontinuities,    while the response of the actually equilibrated system  (two half-sarcomeres in series) is smooth. Note also that the actual response curves do not coincide with either of the two 'constitutive models' and exhibit some intermediate behavior with features mimicking both models simultaneously. The observed discrepancy  is due to the fact that in   a fully equilibrated system none of the contractile elements  is loaded in either soft or hard device and that the overal response of the system is fundamentally non-affine, see also \cite{VILFAN_2003SM,CT_ROPPSM}.

\bibliographystyle{apsrev4-1}

\end{document}